\documentclass{ws-mplb}

\begin{document}

\markboth{Jian-Guo Liu}{Optimization of robustness of scale-free
network to random and targeted attacks}

%
\catchline{}{}{}{}{}
%

\title{OPTIMIZATION OF ROBUSTNESS OF SCALE-FREE
NETWORK TO RANDOM AND TARGETED ATTACKS
}

\author{\footnotesize JIAN-GUO LIU
}

\address{Institute of System Engineering, Dalian University of
Technology, 2 Ling Gong Rd.,\\ Dalian 116024, P R China\\
liujg004@tom.com}

\author{ZHONG-TUO WANG}

\address{Institute of System Engineering, Dalian University of
Technology, 2 Ling Gong Rd., Dalian 116024, P R China\\
wangzt@dlut.edu.cn}

\author{YAN-ZHONG DANG}

\address{Institute of System Engineering, Dalian University of
Technology, 2 Ling Gong Rd.,\\ Dalian 116024, P R China\\
yzhdang@dlut.edu.cn} \maketitle

\begin{history}
\received{(8 February 2005)} \revised{(12 May 2005)}
\end{history}

\begin{abstract}
The scale-fee networks, having connectivity distribution $P(k)\sim
k^{-\alpha}$ (where $k$ is the site connectivity), is very
resilient to random failures but fragile to intentional attack.
The purpose of this paper is to find the network design guideline
which can make the robustness of the network to both random
failures and intentional attack maximum while keeping the average
connectivity $<k>$ per node constant. We find that when $<k>=3$
the robustness of the scale-free networks reach its maximum value
if the minimal connectivity $m=1$ , but when  $<k>$ is larger than
four, the networks will become more robust to random failures and
targeted attacks as the minimal connectivity $m$ gets larger.
\end{abstract}


\section{Introduction}

Recently, there has much interest in the resilience of scale-free
network to random attacks or to intentional attacks on the highest
degree nodes.$^{1-4}$ Many real-world networks are scale-free and
robust to random attacks but vulnerable to intentional attacks. It
is important for us to know the optimal scale free network
guideline to design networks which are optimally robust against
both types of attacks. Although many papers have designed the
optimal network topology, such as the two-peak and three-peak
optimal complex network \cite{13}, but we can not convert its
topology to the theoretical optimization directly disobeying its
evolutionary principle. On the contrary, we should study the
optimal scale-free network guideline to improve the existed
scale-free network robustness.

Studies to data \cite{12,13} have been considered only the case in
which there was only one type of attack in a given network, that
is, the network was subject to either a random attack or a
targeted attack but not subject to different types of attack
simultaneously. A more realistic model is the one in which a
network is subjected to simultaneous targeted and random attacks.
We focus on the network state after intentional and random attacks
which remove fractions $p^{\rm target}$ and $p^{\rm rand}$ of the
original nodes, respectively.

In the scale-free networks, the degree distribution $P(k)$ is the
probability of a node have $k$ connections to other nodes,
typically decreases as a power of $k$. Thus with a fraction $p$ of
the nodes and their connections of the scale-free network are
removed randomly, the random chosen node would have a low degree,
so its removal has little effect on the network. Removal of a
highly connected node could produce a large effect. However, since
such a node may hold significant fractions of the network together
by providing connections between many other nodes. Cohen {\textit
{et. al}} \cite{9} presented a criterion to calculate the
percolation critical threshold of randomly connected networks. If
we attack the scale-free network intentionally: the removal of
sites is not random, but rather sites with the highest
connectivity are targeted first. The numerical simulations suggest
that scale-free network are highly sensitive to this kind of
attack \cite{11}. Cohen {\textit {et. al}} \cite{10} studied the
exact value of the critical fraction needed for disruption and the
size. Thus networks with a given degree distribution may be very
resilient to one type of attack but not to another. This raises
two questions we addressed in this paper: How can we optimize
scale-free network to both random failure and targeted attack and
how to improve the scale free network robustness when the network
size becomes larger. If we construct and maintain a network with
given number of nodes as being proportional to the average number
of links $<k>$ per node in the network. Then, our goal becomes how
to maximize the robustness of a network with $N$ nodes to both
random failure and intentional attack. In our analysis, we compare
the robustness of networks which have the same ``links" of
construction, where we define the cost to be proportional to the
average degree $<k>$ of all the nodes in the network.

This paper is organized as follows. In the first phase, we give
the optimization method to construct the network which is more
robust to random failures. In the second phase, we analysis the
intentional attack to the scale-free network. In the third phase,
we give the optimal strategy of the network design to both random
failure and intentional attack.

\section{Optimal Strategy for Random Failures}

Cohen have studied the properties of the percolation phase
transition in scale-free random networks, and applied a general
criterion for the existence of a spanning cluster:
\begin{equation}\label{F21.1}
\kappa \equiv \frac{<k^2>}{<k>}=2.
\end{equation}
When a fraction $p$ of the nodes are randomly removed, or a
fraction $p$ of the links are randomly removed, the distribution
of site connectivity is changed from the original $P(k)$ to a new
distribution $\widetilde{P}(k)$
\begin{equation}\label{F21.2}
\widetilde{P}(k)=\sum_{k_0\geq k}^KP(k_0)\left(\begin{array}{c}k_0
\\ k
\end{array}\right)(1-p)^kp^{k_0-k}.
\end{equation}
Thus, the critical threshold $p_c$ can be expressed as:
$\frac{<k_0^2>}{<k_0>}(1-p_c)+p_c=2$, that is
$p_c=1-\frac{1}{\kappa_0-1}$, where $\kappa_0\equiv<k_0^2>/<k_0>$
is calculated from the original connectivity distribution. A wide
range of networks have power-law degree distribution:
$P(k)=ck^{-\alpha}, \ \ k=m, m+1, \ldots, K$, where $k=m$ is the
minimal connectivity and $k=K$ is an effective connectivity cutoff
presented in finite networks. The parameter $c$ of the power-law
distribution can be approximate estimated by $c\approx
m^{\alpha-1}(\alpha-1)$. The average connectivity $<k>$ per node
is
$$
<k>=
\frac{(\alpha-1)}{(\alpha-2)}m[1-N^{-\frac{\alpha-2}{\alpha-1}}].
$$
The numerical results show that the exponent $\alpha$ and the
minimum connectivity $m$ have following two relationships: (i)To a
constant average connectivity $<k>$,  the exponent $\alpha$
increases when the minimum connectivity $m$ increases; (ii) To the
minimum connectivity $m=1$, the exponent $\alpha$ decreases when
the average connectivity $<k>$ of the network increases.

The $\kappa_0$ of the scale-free network can be approximated by
\begin{equation}\label{F21.3}
\kappa_0=
\frac{2-\alpha}{3-\alpha}\frac{[K^{(2-\alpha)}-m^{(2-\alpha)}]}{[K^{(3-\alpha)}-m^{(3-\alpha)}]}.
\end{equation}
 Our goal is to maximize the
threshold for random removal with the condition that the average
degree $<k>$ per node is constant. We construct the following
model.

\begin{equation}\label{F2.8}
  \left\{\begin{array}{rl}
    \max & \{1-\frac{1}{\kappa_0-1}\}
    \\[10pt]
     {\rm s.t.}
    &\frac{(\alpha-1)}{(\alpha-2)}m[1-N^{-\frac{\alpha-2}{\alpha-1}}]=<k>,\\[10pt]
    & m\in Z^{+},
  \end{array}
  \right.
\end{equation}
where
$\kappa_0=\frac{2-\alpha}{3-\alpha}\frac{[K^{(2-\alpha)}-m^{(2-\alpha)}]}{[K^{(3-\alpha)}-m^{(3-\alpha)}]}$.
The numerical results suggest that whether the network size $N$ is
very large or not, $p_c$ reaches its maximum value when $m=1$. We
can get the following three conclusions:

\begin{romanlist}[(3)]
\item If the average connectivity $<k>$ per node and the exponent
$\alpha$ of the scale-free network is constant, the robustness of
the network will decrease when the network size becomes larger.
\item If the network size $N$ is constant, the robustness of the
network increases when the average connectivity $<k>$ becomes
larger. \item To the random failures, we have to take several
times cost to increase the robustness of the scale-free network
one percent.
\end{romanlist}

%
%

\section{Breakdown under Intentional Attack}

 Consider the intentional attack, or
sabotage, to the scale-free network, whereby a fraction $p$ of the
sites with the highest connectivity is removed, and the links
emanating from the sites are removed as well. This would make the
cutoff connectivity $K$ of the network reduce to some new value,
$\widetilde{K}<K$. Because the upper cutoff $K$ before intentional
attack can be estimated from
$\sum_{k=K}^{\infty}P(k)=\frac{1}{N}$, the new cutoff
$\widetilde{K}$, after the attack, can be estimated by
\begin{equation}\label{F3.1}
\sum_{k=\widetilde{K}}^{K}P(k)=\sum_{k=\widetilde{K}}^{\infty}P(k)-\frac{1}{N}=p.
\end{equation}
Because $c\approx m^{\alpha-1}(\alpha-1)$, then we have
\begin{equation}\label{F3.2}
p\approx (\frac{\widetilde{K}}{m})^{(\alpha-1)}-\frac{1}{N}.
\end{equation}
If the network size $N$ is very large, $N^{-1}\sim o(10^{-2})$,
the threshold $p$ for intentional attack may be estimated by
$(\widetilde{K}/m)^{(1-\alpha)}=p$.

The intentional attack process to the scale-free network can be
described as following two steps:
\begin{romanlist}[(b)]
\item Removal of the highest connectivity nodes;

\item Removal of the links leading to removed nodes.
\end{romanlist}
So intentional attack to the scale-free network can be considered
as a random removal of links which connect the removed sites with
the remaining sites. We define the probability of removing a link
in the sabotage is $\widetilde{p}$ and all links in the network
have the same probability of being deleted. We have known that the
threshold for random removal of nodes for scale-free network is
\begin{equation}\label{F31.1}
    1-p_c^{\rm rand}=\frac{1}{\kappa_0-1}.
\end{equation}
Then the next task is to find the probability $\widetilde{p}$. The
removal of a fraction $p$ of the sites with the highest
connectivity results in a random removal of links from the
remaining sites that had connected the removed sites with the
remaining sites. The probability $\widetilde{p}$ of a link leading
to deleted site equals the ratio of the number of links belonging
to deleted sites to the total number of links
\begin{equation}\label{F3.3}
\widetilde{p}=\sum_{k=\widetilde{K}}^K \frac{kP(k)}{<\kappa_0>},
\end{equation}
where $\kappa_0$ is the initial average connectivity per node.
With the continuous approximation, this yields
\begin{equation}\label{F3.4}
\begin{array}{rcl}
\widetilde{p} & = & \frac{\int_{k=\widetilde{K}}^K
ck^{1-\alpha}dk}{<\kappa_0>},\\[4pt]
              & = &
              \frac{2-\alpha}{c}(K^{2-\alpha}-m^{2-\alpha})^{-1}\frac{c}{2-\alpha}(K^{2-\alpha}-\widetilde{K}^{(2-\alpha)})\\[4pt]
              & = &
              (K^{2-\alpha}-\widetilde{K}^{(2-\alpha)})(K^{2-\alpha}-m^{2-\alpha})^{-1}\\[4pt]
              & =
              &(\widetilde{K}/m)^{2-\alpha}[1-(\frac{K}{\widetilde{K}})^{2-\alpha}][1-(\frac{K}{m})^{2-\alpha}]^{-1}.\\[4pt]
\end{array}
\end{equation}
In the scale-free network $K>>m$, this yields
\begin{equation}\label{F3.5}
\widetilde{p}=(\widetilde{K}/m)^{2-\alpha}[1-(\frac{K}{\widetilde{K}})^{2-\alpha}].
\end{equation}
Replacing $p_c$ and $K$ in (\ref{F31.1}) with (\ref{F3.5}) and
$\widetilde{K}$, this yields the equation:
\begin{equation}\label{F3.15}
\Big\{1-(\frac{\widetilde{K}}{m})^{(2-\alpha)}[1-(\frac{K}{\widetilde{K}})^{(2-\alpha)}]\Big\}
\Big\{\frac{2-\alpha}{3-\alpha}\frac{\widetilde{K}^{(3-\alpha)}-m^{(3-\alpha)}}{\widetilde{K}^{(2-\alpha)}-m^{(2-\alpha)}}-1\Big\}=1,
\end{equation}
which can be solved numerically to obtain
$\widetilde{K}(m,\alpha,K)$, and then $p_c(m, \alpha)$ can be
retrieved from (\ref{F3.2}). When $N=10^6$, the numerical results
are as follows.

\begin{table}[pt]
\tbl{The value of $p^{\rm target}$ when $N=10^6$ to different
$<k>$.} {\begin{tabular}{@{}c|ccccccc@{}} \toprule & $m=1$  & $m=2$    & $m=3$   &  $m=4$     &  $m=5$ & $m=6$\\
\colrule
$k=3$ & 0.0562 &  0.2010   &   --    &  --        &  --       &  --     \\ 
     $k=4$ & 0.0627 &  0.2152   & 0.4627  &  --        &  --       &  --     \\ 
     $k=5$ & 0.0617 &  0.2100   & 0.4301  &  0.6240    &  --       &  --     \\ 
     $k=6$ & 0.0582 &  0.2003   & 0.4022  &  0.5879    & 0.7172    & --      \\ 
     $k=7$ & 0.0523 &  0.1895   & 0.3780  &  0.5567    & 0.6879    &0.7754   \\ \botrule
\end{tabular} }
\end{table}

From table 1, we can see that: (1) the scale-free network is very
fragile when the minimum connectivity $m$ equals to 1. If
intentionally remove about five percent nodes which have the
highest connectivity of the scale-free network, the network would
collapse. (2) The robustness of the network would increases
dramatically when the minimum connectivity $m$ increases.

The maximum value of $p_c^{\rm target}$ is obtained in the
situation in which all the nodes have the same degree, in which
case the targeted attack becomes equivalent to random failure.
Thus we can use the equation (\ref{F31.1}) to find $p_c^{\rm
target}$. The upper bound is therefore given by
$$
   p_c^{\rm target}\leq 1-\frac{1}{<k>-1}.
$$

\section{Optimization of Robustness of Complex Network}

When the scale free network was attacks randomly and targeted
simultaneously, a metric we can use to measure the robustness of
the network to both random failure and targeted attack is the sum
\begin{equation}\label{F4.1}
p_c^{\rm total}=p_c^{\rm rand}+p_c^{\rm target}.
\end{equation}
This is only one of a number of possible metrics we can use. The
results are, in general, not dependent on the metric chosen.

Our purpose can be stated as follows: for a network of a given
number of nodes $N$, how to maximize $p_c^{\rm total}$ while keep
the number of links constant.

\begin{figure}[th]
\centerline{\psfig{file=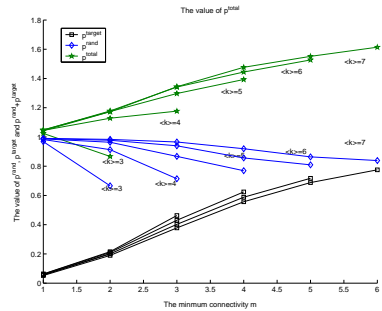,width=5cm}}
\vspace*{8pt} \caption{Random, targeted and total critical
percolation thresholds for scale-free networks to different $<k>$}
\end{figure}

In figure 1, we plot the values of $p_c^{\rm target}$, $p_c^{\rm
rand}$ and $p_c^{\rm total}$ for the minimum connectivity $m$.
Because the exponent $\alpha$ would increase as the minimum
connectivity increase, we can plot the figure for a range of
exponent $\alpha$ too. In the figure, we set the number of nodes
$N=10^6$ and $k=3, 4, 5, 6, 7$. For these choices of $<k>$, we
find that:
\begin{description}
\item[(1)] If $m=1$, or $\alpha$ is around 2.5, $p_c^{\rm total}$
is optimized if the average connectivity $<k>=3$. The telephone
call graph$^{14, 15}$ belongs among this type of network.

\item[(2)] If $<k>$ is larger than four, the network would become
more robust as the minimum connectivity (or the exponent $\alpha$)
increases. Many networks belong among this type of network, such
as coauthorship network. $^{16-21}$ This conclusion shows that the
relation between the minimum connectivity nodes of the network is
very important. More tight the relationship between the minimum
degree nodes of the scale-free networks, more robust the network
would be.
\end{description}

\section{Discussion and Summary}

 Although the theoretical optimal networks to both random and
targeted attacks have been designed, it is very important that
some large size networks, such as the internet, is a
self-organizing system, evolve and drastically changes over time
according to evolutionary principle dictated by the interplay
between cooperation and competition. To the exist growing
scale-free network, we can not convert its topology into the
theoretical optimization directly disobeying its evolutionary
principle. But we can improve the network robustness. Firstly, we
must know the optimal guideline to maximum the scale-free network
robustness.

The $p_c^{\rm total}$ characterizes the robustness of the
scale-free network to random failures and intentional attacks
simultaneously. In this paper we analyze the scale-free network
robustness to both random failures and intentional attacks
simultaneously and find that the minimum degree of the network is
very important to the scale-free network robustness with a
constant $<k>$. To an exist growing scale-free network, if the
network average degree per node is around three, we should keep
its minimum degree around one to improve the network robustness.
But if the network average connectivity is larger than four, we
should add its minimum degree to improve its robustness to both
attacks, which mean that the relationship between the minimum
connectivity nodes would become very important for the network
robustness.




%

\section*{Acknowledgments}

This research was supported by the Chinese Natural Science
Foundation (Grant No. 70431001, 70271046).


\end{document}